\documentstyle[12pt]{article}
\newcommand{\Ir}{Z\!\!\!Z}
\newcommand{\idty}{{\leavevmode{\rm 1\mkern -5.4mu I}}}
\newcommand{\Ibb}[1]{ {\rm I\ifmmode\mkern -3.6mu\else\kern -.2em\fi#1}}
\newcommand{\ibb}[1]{\leavevmode\hbox{\kern.3em\vrule
     height 1.2ex depth -.3ex width .2pt\kern-.3em\rm#1}}
\newcommand{\Cx}{{\ibb C}}
\newcommand{\Rl}{{\Ibb R}}
\normalbaselines
\begin{document}
\count0 = 1

\begin{tabbing}
\hspace*{12cm} \= GOET-TP 97/93   \\
               \> December 1993
\end{tabbing}
\vskip.5cm

\begin{center}
\vspace*{1.0cm}
{\LARGE
{\bf Differential Calculi on Quantum \\[1mm]
(Sub-) Groups and Their
Classical Limit}}

\vskip 1.5cm

{\large
{\bf  Folkert M\"uller-Hoissen}}

\vskip 0.5 cm
Institut f\"ur Theoretische Physik, D-37073
            G\"ottingen, Germany

\end{center}

\vspace{1 cm}

\begin{abstract}
For the two-parameter matrix quantum group $GL_{p,q}(2)$ all
bicovariant differential calculi (with a four-dimensional space
of 1-forms) are known. They form a one-parameter family. Here,
we give an improved presentation of previous results by using a
different parametrization. We also discuss different ways to
obtain bicovariant calculi on the quantum subgroup $SL_q(2)$.
For those calculi, we do {\em not} obtain the ordinary differential
calculus on $SL(2)$ in the classical limit. The structure which emerges
here can be generalized to a nonstandard differential calculus on an
arbitrary differentiable manifold and exhibits relations with
stochastic calculus and `proper time' relativistic quantum
theories.

\end{abstract}
\vskip1cm

\begin{center}
\begin{minipage}[t]{10cm}
To appear in the proceedings of the International Symposium
on ``Generalized Symmetries in Physics'', ASI Clausthal, July 1993.
\end{minipage}
\end{center}

\newpage

\renewcommand{\theequation} {\arabic{section}.\arabic{equation}}
\section{Introduction}
\setcounter{equation}{0}
Differential geometry of Lie groups plays an important role
in the mathematical modelling of physical theories. In
particular, this is the case for classical gauge theories
formulated in terms of connections on principal fiber bundles.
Since a Hopf algebra or quantum group can be regarded
as a generalization of the notion of a group, it is tempting
to also generalize the corresponding notions of differential
geometry (see \cite{Brze+Maji92-gauge}, in particular). Besides
promising mathematical aspects of such
a generalization, there is a hope to obtain interesting
`deformations' of physical models, like the gauge theory models
of elementary particle physics.

More generally, a notion of differential forms has
been introduced for an arbitrary associative algebra $\cal A$
\cite{Conn86}. One can enlarge $\cal A$ to a {\em differential
algebra}. This is a $\Ir$-graded associative algebra
$\bigwedge({\cal A}) = \bigoplus_{r \geq 0} \bigwedge^r({\cal A})$
where $\bigwedge^0 = \cal A$ and the spaces $\bigwedge^r({\cal A})$
of {\em $r$-forms} are generated as $\cal A$-bimodules via the action
of an {\em exterior derivative} $\mbox{d} \, : \, \bigwedge^r({\cal A})
\rightarrow \bigwedge^{r+1}({\cal A})$. The latter is a linear operator
acting in such a way that $\mbox{d}^2 =0$ and $\mbox{d}(\omega \omega')
= (\mbox{d} \omega) \, \omega' + (-1)^r \omega \, \mbox{d} \omega'$
where $\omega$ and $\omega'$ are $r$- and $r'$-forms, respectively.
There are many differential algebras associated with an algebra
$\cal A$. But all of them can be obtained as a quotient of a
maximal differential algebra, the {\em universal differential
envelope}, by some ideal. In particular, if
$\cal A$ is the algebra of polynomials of $n$ independent elements,
we might want the associated space of 1-forms to be
$n$-dimensional as a left (or right) $\cal A$-module. This does not
restrict the possible differential algebras very much, however.
In general, there seems not to be a kind of functorial way to
associate such a differential algebra with a given algebra $\cal A$.
On the other hand, it turned out that different choices of differential
algebras are actually of interest from a mathematical and physical
point of view (cf the examples in
\cite{DMHS93-latt,DMH92-grav,DMH93-stoch,MH92-habil}).

For the case of matrix quantum groups, Woronowicz introduced the
notions of left-, right- and bi-covariant differential calculus
\cite{Woro89}. Bicovariance was soon accepted as the most natural
condition for a differential algebra. Woronowicz gave two examples
of bicovariant differential algebras on $SU_q(2)$ (the socalled
$4D_\pm$ calculi) \cite{Woro89}. At that time, it was not known
how many bicovariant differential calculi exist on $SU_q(2)$ (and
other quantum groups). Later, it turned out that Woronowicz
already found {\em all} bicovariant calculi on $SU_q(2)$
\cite{Stac92,MH+Reut93}. In the meantime, a large number of papers
appeared dealing with examples of bicovariant differential calculi
on special (classes of) quantum groups (see \cite{MH+Reut93} for an
extensive list of references). However, according to our knowledge
there are only few papers which go beyond examples and give a complete
description of the possible bicovariant differential calculi on
certain quantum groups \cite{Stac92,MH92,Schm+Schu93}.
In \cite{MH92} {\em all} bicovariant differential calculi
on the two-parameter quantum group $GL_{p,q}(2)$ were found (see
also \cite{MH92-habil}).
They form a family which depends on an additional parameter
$s$ ($q$, $p$ and $s$ are complex numbers). In section 2, this
family and its classical limit is described using a simplifying
parametrization which greatly improves the presentation in
\cite{MH+Reut93,MH92}. In section 3, we consider two
ways in which the family of bicovariant calculi on $GL_{p,q}(2)$
induces corresponding calculi on the quantum
subgroup $SL_q(2)$. Also the classical limit of bicovariant calculus
on $SL_q(2)$ is dicussed. Here we are led to a generalization
\cite{DMH92-grav,DMH93-stoch,MH92-habil} of the ordinary calculus of
differential forms on a manifold on which we comment in section 4.

\section{Bicovariant differential calculi on $GL_{p,q}(2)$}
\setcounter{equation}{0}
{\bf 2.1 The quantum general linear group in two dimensions.}
We recall that the quantum group $GL_{p,q}(2)$ is the Hopf algebra
$\cal A$ generated by $a,b,c,d$ and the unit $\idty$, satisfying the
commutation relations
\begin{eqnarray}
  \begin{array}{l@{\:=\:}l@{\qquad}l@{\:=\:}l@{\qquad}l@{\:=\:}l}
 a\,c & q \,c\,a &  b\,d & q \,d\,b &
 a\,d & d\,a + q\,c\,b-(1/q)\,b\,c  \\
 b\,c & (q/p)\,c\,b &  a\,b & p\,b\,a &
 c\,d & p\,d\,c
\end{array}        \label{GL-crs}
\end{eqnarray}
where $p,q \in \Cx \setminus \lbrace 0 \rbrace$. The existence of an
inverse ${\cal D}^{-1}$ of the `quantum determinant'
$ {\cal D} := a d-p \, b c $ is required.
The {\em coproduct} is the homomorphism determined by
\footnote{Here and in the following we use a compact notation for
$\Delta(a) = a \otimes a + b \otimes c$ etc.}
\begin{eqnarray}
 \Delta \left(\begin{array}{cc} a & b\\ c & d \end{array}\right)
 = \left(\begin{array}{cc} a & b\\ c & d \end{array}\right)
   \dot{\otimes}
   \left(\begin{array}{cc} a & b\\ c & d \end{array}\right)
 := \left(\begin{array}{cc} a\otimes a+b\otimes c & a\otimes b
   +b\otimes d  \\
  c\otimes a+d\otimes c & c\otimes b+d\otimes d \end{array}\right)
\end{eqnarray}
and the {\em antipode} is the anti-homomorphism
$\, S \, : \, {\cal A} \, \rightarrow \, {\cal A} \,$  with
\begin{eqnarray}
 S \left(\begin{array}{cc} a & b  \\ c & d
   \end{array}\right) =
   {\cal D}^{-1}\,\left(\begin{array}{cc} d & -b/q \\ -q\,c & a
   \end{array}\right)  \; .
\end{eqnarray}
In addition, $\Delta(\idty) = \idty \otimes \idty$ and $S(\idty)=\idty$.
\vskip.3cm
\noindent
{\bf 2.2 Bicovariant differential calculus.}
The central object of {\em first order} differential calculus is the
exterior derivative $\mbox{d} \, : \, {\cal A} \,\rightarrow \,
\Lambda^1({\cal A})$ satisfying the Leibniz rule
$ \mbox{d} (f h) = (\mbox{d} f)\,h + f\,\mbox{d} h \; (\forall\,
  f,h \in {\cal A})$.
The space of 1-forms $\Lambda^1({\cal A})$ is generated as an
$\cal A$-bimodule by the differentials of $a,b,c,d$. It is
furthermore required that the differentials of $a,b,c,d$ form a basis
of $\Lambda^1({\cal A})$ as a left $\cal A$-module.
In order to achieve this, one has to find commutation relations between
$a,b,c,d$ and their differentials which are consistent with the
differential algebra structure.
\vskip.2cm

A {\em left-coaction}  $ \, \Delta_{\cal L} \, : \; \Lambda^1({\cal A})
\, \rightarrow \, {\cal A} \otimes \Lambda^1({\cal A}) \, $
extends $\Delta$ as a bimodule homomorphism to 1-forms such that
\begin{eqnarray}
 \Delta_{\cal L}\,\left(\begin{array}{cc} \mbox{d} a & \mbox{d} b \\
                                   \mbox{d} c & \mbox{d} d
                 \end{array} \right)
 &=& \left(\begin{array}{cc} a & b  \\
                             c & d
                 \end{array} \right)
     \dot{\otimes}
     \left(\begin{array}{cc} \mbox{d} a & \mbox{d} b \\
                                   \mbox{d} c & \mbox{d} d
                 \end{array} \right)    \; .
\end{eqnarray}
In the same way a {\em right-coaction} $\, \Delta_{\cal R}\, : \;
\Lambda^1({\cal A})\,\rightarrow \, \Lambda^1({\cal A})\otimes {\cal A}
\, $ is a bimodule homomorphism with
\begin{eqnarray}
 \Delta_{\cal R}\,\left(\begin{array}{cc} \mbox{d} a & \mbox{d} b \\
                                   \mbox{d} c & \mbox{d} d
                 \end{array} \right)
 = \left(\begin{array}{cc} \mbox{d} a & \mbox{d} b \\
                             \mbox{d} c & \mbox{d} d
                 \end{array} \right)
    \dot{\otimes}
    \left(\begin{array}{cc} a & b \\
                             c & d
                 \end{array} \right)   \; .
\end{eqnarray}
If $\Delta_{\cal L}$ and $\Delta_{\cal R}$ exist, the (first order)
differential calculus is called {\em bicovariant} \cite{Woro89}.
\vskip.2cm

Assuming the existence of $\Delta_{\cal L}$, there is a basis of
(left-coinvariant) {\em Maurer-Cartan 1-forms} $\theta^K$ in
$\Lambda^1({\cal A})$ given by
\begin{eqnarray}
 \left(\begin{array}{cc} \theta^1 & \theta^2 \\
                             \theta^3 & \theta^4
                 \end{array} \right)
 = S\,\left(\begin{array}{cc} a & b \\
                             c & d
                 \end{array} \right)  \;
   \mbox{d} \, \left(\begin{array}{cc} a & b \\
                                       c & d
                     \end{array} \right) \; .
\end{eqnarray}
Commutation relations between the generators of
$\cal A$ and their differentials can be expressed
in terms of the Maurer-Cartan 1-forms,
\begin{eqnarray}   \label{theta-f}
 \theta^K\,f = \Theta(f)^K_L\,\theta^L \qquad (\forall f\in {\cal A})
               \; .
\end{eqnarray}
Compatibility with $\Delta_{\cal L}$ leads to
\begin{eqnarray}
 \Theta\,\left(\begin{array}{cc} a & b \\ c & d \end{array}\right)
 = \left(\begin{array}{cc} a & b \\ c & d \end{array}\right)\,
   \left(\begin{array}{cc} A & B \\ C & D \end{array}\right)
                  \label{Theta-ABCD}
\end{eqnarray}
where $A,B,C,D$ are $4\times 4$-matrices (with complex entries).
Associativity of $\Lambda^1({\cal A})$ and (\ref{theta-f}) require
$\Theta(f h) = \Theta(f) \Theta(h)$ which means that
$A,B,C,D$ have to form a representation of $a,b,c,d$.
(\ref{theta-f}) and (\ref{Theta-ABCD}) imply
\begin{eqnarray}
  \theta^K \, a = (a \, A^K_L + b \, C^K_L)  \, \theta^L
       \quad , \quad
  \theta^K \, b = (a \, B^K_L + b \, D^K_L)  \, \theta^L
\end{eqnarray}
and the corresponding relations with $a$ replaced by $c$ and $b$
replaced by $d$. The consistency conditions for first order
bicovariant differential calculus \cite{Woro89} were completely solved
for $GL_{p,q}(2)$ in \cite{MH92} using computer algebra (see also
\cite{MH93}). We found that there is a one-parameter set of such
calculi.

\vskip.3cm \noindent
{\em Theorem} \cite{MH92} \\
Let $r := p \,q \neq 0,\pm 1$ and $t \in \Cx$, $t \neq 0$ and
$t \neq r(1+r)/(1+r^2)$.
\footnote{The last condition ensures that $N \neq 0$. In the case
excluded by this condition, there are no calculi when $r \neq \pm 1$
\cite{MH+Reut93}. $t=0$ has to be excluded because in that case one
finds $\theta^K \, {\cal D} = 0$ which conflicts with the existence
of ${\cal D}^{-1}$. We may admit $r=1$ in the theorem, but in that case
additional calculi exist \cite{MH92}.}
All bicovariant first order differential calculi on $GL_{p,q}(2)$
-- for which the differentials of $a,b,c,d$ form a basis of
$\Lambda^1({\cal A})$ as a left $\cal A$-module --
are given by \footnote{Here we use a different parametrization
as in \cite{MH+Reut93,MH92}. The reason is that writing
$A^1_1 = t \, (1+r)/r + r \, A^1_4 -1$ with a new parameter $t$, the
quadratic relation between $A^1_1$ and $A^1_4$ which was obtained in
\cite{MH92} (equation $(6.25)$ therein) simply becomes the expression
for $A^1_4$ in (\ref{dc-entries}). $A^1_4$ is the complex parameter
$s$ in \cite{MH+Reut93,MH92}. $t$ is the parameter $s$ in
\cite{Asch+Cast93}.}
\begin{eqnarray}
 \begin{array}{ll}
 A =   \left(\begin{array}{cccc}
       A^1_1 & 0 & 0 & A^1_4     \\
       0 & t/p & 0 & 0  \\
       0 & 0 & t/q & 0  \\
       A^4_1 & 0 & 0 & 1 - r \, A^1_4 \end{array}\right)      &
 B =   \left(\begin{array}{cccc}
       0     & t-1   & 0 & 0     \\
       0     & 0     & 0 & 0         \\
       B^3_1 & 0     & 0 & B^3_4 \\
       0     & t/r-1 & 0 & 0     \end{array}\right)      \\
                                                     \\
 C =   \left(\begin{array}{cccc}
       0 & 0 & t-1 & 0     \\
       (q/p)B^3_1 & 0 & 0 & (q/p) B^3_4  \\
       0 & 0 & 0 & 0  \\
       0 & 0 & t/r-1 & 0 \end{array}\right)           &
 D =   \left(\begin{array}{cccc}
       1-A^1_4 & 0 & 0 & D^1_4 \\
       0 & t/p & 0 & 0  \\
       0 & 0 & t/q & 0  \\
       r \, A^1_4 & 0 & 0 & D^4_4 \end{array}\right)
  \end{array}          \label{GLpq-mat}
\end{eqnarray}
where
\begin{eqnarray}
 A^1_1 &=& \lbrack r^2 (r \, t -1)(t-1) + r \, t \, (t-2) + t^2
           \rbrack / (r^3 \, N)                         \nonumber  \\
 A^1_4 &=& (r-t)(t-1) / (r^2 \, N)                      \nonumber  \\
 A^4_1 &=& (t-r)(r^2 t -r t -r + t)/(r^3 \, N)          \nonumber  \\
 B^3_1 &=& t \, (r-t) (r-1)/(q \, r^2 \, N)             \nonumber  \\
 B^3_4 &=& t \, p \, (r-1) (t-1)/(r^2 \, N)             \nonumber  \\
 D^1_4 &=& (t-1) (r^2 t - r^2 -r \, t + t)/(r^2 \, N)   \nonumber  \\
 D^4_4 &=& \lbrack r^3 (t-1)^2 + r^2 t (t-1) - r \, t + t^2 \rbrack
           / (r^3 \, N)                          \label{dc-entries}
\end{eqnarray}
with  $N := \lbrack t \, (1+r^2) - r \, (1+r) \rbrack / r^2 $.
                                               \hfill {\Large $\Box$}
\vskip.3cm
In terms of the differentials, the commutation relations
(\ref{theta-f}) for the bicovariant differential calculi
are not quadratic relations if $t \neq 1,r$. For example,
\begin{eqnarray}
\mbox{d}a \, a &=& (A^1_1+ A^1_4 \, p \, r^2 \, {\cal D}^{-1}
           \, b \, c) \, a \, \mbox{d}a
           - A^1_4 \, {\cal D}^{-1}  \, a^2 \, (q \, c \, \mbox{d}b
           + p \, b \, \mbox{d}c - a \, \mbox{d}d) \; .
                                 \label{da-a}
\end{eqnarray}
The differential of an element $f \in {\cal A}$ can be expressed as
\cite{Woro89,MH92}
\begin{eqnarray}
  \mbox{d} f = {1\over N} \, \lbrack \, \vartheta \, , \, f \, \rbrack
                                          \label{df-vartheta}
\end{eqnarray}
where $N$ is defined in the theorem and
\begin{eqnarray}
  \vartheta := \theta^1 + {1 \over r} \, \theta^4
                                           \label{vartheta}
\end{eqnarray}
is a bi-coinvariant 1-form.
Bicovariant first order differential calculi always admit an
extension to higher orders \cite{Woro89}. Differential forms of higher
order are obtained by applying $\mbox{d}$ to 1-forms (and then also
higher forms) using $\mbox{d}^2 = 0$ and the graded Leibniz rule.
Bicovariance guarantees that there are commutation relations between
the 1-forms which are compatible with these structures.
(\ref{df-vartheta}) then holds more generally with $f$ replaced by
any form if the commutator is replaced by a graded commutator
\cite{Woro89}. We refer to \cite{MH92} for the corresponding results
in the case of $GL_{p,q}(2)$.
\vskip.3cm
\noindent
{\bf 2.3 An $R$-matrix formulation.} In terms of the new basis of
left-coinvariant 1-forms
\begin{eqnarray}
  \begin{array}{c@{\:=\:}l@{\qquad}c@{\:=\:}l}
 \omega^1{}_1 & (p / r^2 \, N \, t) \, \lbrack (r-t) \, \theta^1
            + r \, (t-1) \, \theta^4 \rbrack
              & \omega^1{}_2 & - (p / q \, t) \, \theta^2   \\
 \omega^2{}_2 & (p / r^2 \, N \, t) \, \lbrack ( t (r^2-r+1) -r ) \,
            \theta^1 + (r-t) \, \theta^4 \rbrack
          & \omega^2{}_1 & - (1 / t) \, \theta^3  \; ,
  \end{array}
\end{eqnarray}
the commutation relations with elements of $\cal A$ are given by
\begin{eqnarray}
  \begin{array}{c@{\:=\:}l@{\quad\quad}c@{\:=\:}l}
 \omega^1{}_1 \, a & (t/r) \, a \, \omega^1{}_1
                   & \omega^1{}_1 \, b & t \, b \, \omega^1{}_1    \\
 \omega^1{}_2 \, a & t \, \lbrack p^{-1} \, a \, \omega^1{}_2 + r^{-1}
                 (1-r) \, b \, \omega^1{}_1 \rbrack
                   & \omega^1{}_2 \, b & (t/p) \, b \, \omega^1{}_2 \\
 \omega^2{}_1 \, a & (t/q) \, a \, \omega^2{}_1
                   & \omega^2{}_1 \, b & t \, \lbrack q^{-1} \, b \,
                     \omega^2{}_1 + r^{-1} (1-r) \, a \, \omega^1{}_1
                     \rbrack     \\
 \omega^2{}_2 \, a & t \, \lbrack a \, \omega^2{}_2 + q^{-1} (1-r)
                     \, b \, \omega^2{}_1 \rbrack
                   & \omega^2{}_2 \, b & \begin{minipage}[t]{5cm}
                     $ (t/r) \, \lbrack  b \, \omega^2{}_2
                       + (r-1)^2 \, b \, \omega^1{}_1 $ \\
                 $ + q \, (1-r) \, a \, \omega^1{}_2 \rbrack \; .$
                 \end{minipage}
  \end{array}                         \label{omega-a,b}
\end{eqnarray}
For $p=q$, these relations can be found in \cite{Asch+Cast93}. In terms
of the 1-forms $\omega^i{}_j, \, i,j = 1,2$, the commutation relations
look much simpler than the corresponding relations with $\theta^K$. In
particular, the parameter $t$ only appears as a common factor on the
right hand sides of (\ref{omega-a,b}). However, one has to keep
in mind that the 1-forms $\omega^i{}_j$ -- when expressed in terms of
the differentials of $a,b,c,d$ or the Maurer-Cartan 1-forms -- depend on
$t$ (and $p,q$) in a rather complicated way.
The relations (\ref{omega-a,b}) can be expressed in terms of the
$R$-matrix of $GL_{p,q}(2)$ as follows\footnote{See also
\cite{Asch+Cast92} for the case of $GL_q(3)$.},
\begin{eqnarray}
 \omega^i{}_j \, T^k{}_\ell = t \, (q/p) \, T^k{}_m \,
 (R^{-1})^{mn}{}{}_{uj} \, (R^{-1})^{iu}{}{}_{v \ell} \,
 \omega^v{}_n
\end{eqnarray}
where $T$ is the matrix with entries $a,b,c,d$ and
\begin{eqnarray}
 R^{-1} = \left( \begin{array}{cccc}
          q^{-1} & 0        & 0   & 0 \\
               0 & 1        & 0   & 0 \\
               0 & q^{-1}-p & p/q & 0 \\
               0 & 0        & 0 & q^{-1} \end{array} \right)  \; .
\end{eqnarray}
Rows and columns of the matrix are numbered by
$(1,1),(1,2),(2,1),(2,2)$.
To this expression for the bicovariant differential calculi one
is led by applying the recipe of \cite{Jurc91} (based on the
techniques of \cite{FRT90}) with the slight generalization given
in \cite{Sun+Wang92,Asch+Cast93}). It is interesting that this
procedure already exhausts the possible bicovariant calculi. This also
holds (with a further refinement) for the quantum group $GL_q(3)$ for
which all bicovariant differential calculi have recently been obtained
\cite{Bres93} using the methods of \cite{MH92}. Also in this case we
have a one-parameter family and half of it was already found in
\cite{Asch+Cast92}. This suggests that, more generally, on $GL_q(n)$
($n \geq 2$) the bicovariant calculi form a one-parameter set. There
are indeed partial results \cite{Isae+Pyat93} substantiating this
conjecture.
\vskip.3cm
\noindent
{\bf 2.4 The classical limit.}
In terms of $x^1:=a, \, x^2:=b, \, x^3:=c, \, x^4:=d$, the
commutation relations between $x^\mu$ and $\mbox{d} x^\nu$ (cf
(\ref{da-a}) and \cite{MH+Reut93}) take the following form in the
classical limit $p,q \to 1$,
\begin{eqnarray}       \label{x-dx}
     \lbrack \, x^\mu \, , \, \mbox{d} x^\nu \, \rbrack
     = \tau \, g^{\mu \nu}
\end{eqnarray}
with
\begin{eqnarray}
  \tau &:=& - s \, \vartheta = s \, \lbrack \mbox{d} x^1 \, x^4
           - \mbox{d} x^2 \, x^3 - \mbox{d} x^3 \, x^2 + \mbox{d} x^4
           \, x^1 \rbrack =: \mbox{d} x^\mu \, \tau_\mu
                                  \label{tau}     \\
  g^{\mu \nu} &:=& (x^1 x^4 - x^2 x^3)^{-1} \, x^\mu \, x^\nu
                  + 4 \, \lbrack \delta^{(\mu}_2 \, \delta^{\nu)}_3
                  - \delta^{(\mu}_1 \, \delta^{\nu)}_4 \rbrack
                                   \label{g}
\end{eqnarray}
where indices in brackets are symmetrized. Here we have
\begin{eqnarray}                   \label{s-t}
                      s = (1-t)/2
\end{eqnarray}
for $t \neq 1$ (cf the assumptions in the
theorem) if we regard $t$ as a parameter which does not dependent
on $p$ or $q$ (otherwise the limit will depend on the choice of $t$
as a function of $p$ and $q$, cf section 3.3).
The matrix $g$ is degenerate since $g^{\mu \nu} \, \tau_\nu = 0$
which reminds us of a `Galilei structure' (see also
\cite{DMH93-stoch}, appendix B).
The 1-form $\tau$ commutes with $x^\mu, \, \mu=1, \ldots, 4$,
anticommutes with all 1-forms and satisfies $\mbox{d} \tau = 0$.

\section{From differential calculus on $GL_q(2)$ to differential
         calculus on $SL_{q}(2)$}
\setcounter{equation}{0}
In this section we restrict the deformation parameters of $GL_{p,q}(2)$
by $p=q$. The quantum group is then called $GL_q(2)$. In this case the
quantum determinant $\cal D$ (see section 2.1) becomes central, i.e. it
commutes with all elements of $\cal A$. The condition
\begin{eqnarray}   \label{det=1}
   {\cal D} = a \, d - q \, b \, c = \idty
\end{eqnarray}
then defines the quantum subgroup $SL_q(2)$. In the following it is
shown that there are two different ways to obtain bicovariant
differential calculi on $SL_q(2)$ from the family of bicovariant
calculi on $GL_q(2)$ (see also \cite{MH+Reut93}).
\vskip.3cm
\noindent
{\bf 3.1 The direct way.}
Imposing the condition (\ref{det=1}) on the family of bicovariant
differential algebras on $GL_q(2)$ requires that all 1-forms commute
with $\cal D$. This means that
\begin{eqnarray}
    {\bf 1} = A \, D - q \, B \, C = (t / q)^2 \, {\bf 1}
\end{eqnarray}
(where $\bf 1$ is the $4\times 4$ unit matrix) and restricts the
parameter $t$ to the values
\begin{eqnarray}
               t_{\pm} = \pm q  \; .
\end{eqnarray}
For the general bicovariant calculus on $GL_q(2)$ one finds
\footnote{The factor $\cal D$ is missing on the rhs of (4.7) in
\cite{MH+Reut93}.}
\begin{eqnarray}
  \mbox{d} \, {\cal D} = - {(t-q)(t+q) \over q^2 \, N} \, {\cal D}
                         \; \vartheta
\end{eqnarray}
with $N$ and $\vartheta$ defined in section 2.
Differentiation of (\ref{det=1}) leads to the constraint
$\mbox{d} {\cal D} = 0$ which is identically satisfied when
$t=t_\pm$. Hence, there are two bicovariant differential calculi on
$GL_q(2)$ which are consistent with the constraint (\ref{det=1}):
\vskip.2cm
\noindent
(1) For $t=t_+$ the matrices $A,B,C,D$ take the following form.
\begin{eqnarray}
 \begin{array}{ll}
 A =   \left(\begin{array}{cccc}
       {q^4+q^3+q^2+1 \over q \, (q^2+q+1)} & 0 & 0 & {1 \over
                                                       q^2+q+1}     \\
       0 & 1 & 0 & 0  \\
       0 & 0 & 1 & 0  \\
       -{q^3+q^2-1 \over q \, (q^2+q+1)} & 0 & 0 & {q+1 \over
                                                    q^2+q+1}
                                     \end{array}\right)      &
 B =   \left(\begin{array}{cccc}
       0 & q-1 & 0 & 0     \\
       0 & 0 & 0 & 0         \\
       {q+1 \over q^2+q+1} & 0 & 0 & {q \, (q+1) \over q^2+q+1} \\
       0 & {1\over q}-1 & 0 & 0     \end{array}\right)      \\
                                                     \\
 C =   \left(\begin{array}{cccc}
       0 & 0 & q-1 & 0     \\
       {q+1 \over q^2+q+1} & 0 & 0 & {q \, (q+1) \over q^2+q+1}  \\
       0 & 0 & 0 & 0  \\
       0 & 0 & {1\over q}-1 & 0 \end{array}\right)           &
 D =   \left(\begin{array}{cccc}
       {q \, (q+1) \over q^2+q+1} & 0 & 0 & {q^3-q-1 \over q^2+q+1} \\
       0 & 1 & 0 & 0  \\
       0 & 0 & 1 & 0  \\
       {q^2 \over q^2+q+1} & 0 & 0 & {q^4+q^2+q+1 \over
                                     q \, (q^2+q+1)}
                                        \end{array}\right)
  \end{array}      \label{ABCD-t+}
\end{eqnarray}
\vskip.3cm
\noindent
(2) For $t=t_-$ the matrices $A,B,C,D$ are given by
\begin{eqnarray}
 \begin{array}{ll}
 A =   \left(\begin{array}{cccc}
       -{q^4-q^3+q^2+1 \over q \, (q^2-q+1)} & 0 & 0 & {1 \over
                                                       q^2-q+1}     \\
       0 & -1 & 0 & 0  \\
       0 & 0 & -1 & 0  \\
       -{q^3-q^2+1 \over q \, (q^2-q+1)} & 0 & 0 & {1-q \over
                                                    q^2-q+1}
                                     \end{array}\right)      &
 B =   \left(\begin{array}{cccc}
       0 & -(q+1) & 0 & 0     \\
       0 & 0 & 0 & 0         \\
       {q-1 \over q^2-q+1} & 0 & 0 & {q \, (1-q) \over q^2-q+1} \\
       0 & -(1+{1\over q}) & 0 & 0     \end{array}\right)      \\
                                                     \\
 C =   \left(\begin{array}{cccc}
       0 & 0 & -(q+1) & 0     \\
       {q-1 \over q^2-q+1} & 0 & 0 & {q \, (1-q) \over q^2-q+1}  \\
       0 & 0 & 0 & 0  \\
       0 & 0 & -(1+{1\over q}) & 0 \end{array}\right)           &
 D =   \left(\begin{array}{cccc}
       {q \, (q-1) \over q^2-q+1} & 0 & 0 & -{q^3-q+1 \over q^2-q+1} \\
       0 & -1 & 0 & 0  \\
       0 & 0 & -1 & 0  \\
       {q^2 \over q^2-q+1} & 0 & 0 & -{q^4+q^2-q+1 \over
                                     q \, (q^2-q+1)}
                                        \end{array}\right)
  \end{array}              \label{ABCD-t-}
\end{eqnarray}
\vskip.3cm
\noindent
{\em Theorem} \cite{MH+Reut93}   \\
Let $q \neq 0, \pm 1, \pm i$. The $t_\pm$ calculi (with
$q^2 \pm q +1 \neq 0$) are the only bicovariant differential calculi
on $SL_q(2)$.
                                     \hfill {\Large $\Box$}
\vskip.3cm
\noindent
The two calculi on $SL_q(2)$ induce the $4D_\pm$ calculi
\cite{Woro89} on $SU_q(2)$. The uniqueness of the latter has
been shown in \cite{Stac92}.
\vskip.3cm
\noindent
{\bf 3.2 An indirect way.}
There is another simple way to obtain a differential calculus
on $SL_q(2)$ from a calculus on $GL_q(2)$. For the special differential
calculus with $t=1$ it has been considered in \cite{SchuWZ92}.
Let $T$ denote the matrix with entries $a,b,c,d$ satisfying the
$GL_q(2)$ commutation relations. Furthermore, let us assume that
${\cal D}^{-1/2}$ exists and commutes with all elements of $GL_q(2)$
(note that $\cal D$ is central). Then
\begin{eqnarray}           \label{theta-sq(det)}
 \theta^K \, {\cal D}^{-1/2} = \pm (q / t) \, {\cal D}^{-1/2} \,
   \theta^K
\end{eqnarray}
(for $t \neq 0$).
The entries of
\begin{eqnarray}
 \hat{T} := {\cal D}^{-1/2} \, T =:
    \left( \begin{array}{cc}
    \hat{a} & \hat{b} \\ \hat{c} & \hat{d}
    \end{array} \right)
\end{eqnarray}
satisfy the $GL_q(2)$ commutation relations
and furthermore $\hat{\cal D} = \mbox{det}_q \hat{T} = \idty$. They
generate $SL_q(2)$ as a subalgebra of $GL_q(2)$ and the
differential calculus can be restricted to it. We can introduce
corresponding Maurer-Cartan 1-forms
\begin{eqnarray}
  \left( \begin{array}{cc} \hat{\theta^1} & \hat{\theta^2} \\
                           \hat{\theta^3} & \hat{\theta^4}
         \end{array}   \right)
    := S(\hat{T}) \, \mbox{d} \hat{T}
     =  \pm (q/t) \, \left( \begin{array}{cc} \theta^1 & \theta^2 \\
        \theta^3 & \theta^4   \end{array} \right)
        + {1\over N} \, (\pm q/t-1) \, \left( \begin{array}{cc}
        \vartheta & 0 \\
        0 & \vartheta   \end{array} \right) \; . \label{MC-hat}
\end{eqnarray}
To derive the last expression, we made use of (\ref{df-vartheta}) and
(\ref{theta-sq(det)}).
It allows us to calculate commutation relations between the
1-forms $\hat{\theta}^K$ and the entries of $T$ from the
corresponding commutation relations of a bicovariant differential
calculus on $GL_q(2)$.
In this way, each bicovariant differential calculus on
$GL_q(2)$ induces two corresponding bicovariant differential
calculi on the subalgebra $SL_q(2)$. In accordance with the theorem
in section 3.1, the latter do not dependent on the value of
$t$. More generally, one obtains the following result about the
structure of the bicovariant calculi on $GL_q(2)$.
\vskip.3cm
\noindent
{\em Theorem} \cite{MH+Reut93}  \\
Let $q \neq 0, \pm 1,\pm i$, $t \neq 0$ and $t \neq q^2
(1+q^2)/(1+q^4)$.
In terms of the ($SL_q(2)$ Maurer-Cartan) 1-forms (\ref{MC-hat})
and the algebra elements $\hat{a}, \hat{b}, \hat{c}, \hat{d},
{\cal D}^{1/2}$ {\em all} bicovariant differential calculi on
$GL_q(2)$ are determined by \footnote{The parameter $t$ only enters
the last relation. It appears, however, implicitly in the relation
between the $GL_q(2)$ and the $SL_q(2)$ Maurer-Cartan forms.}
\begin{eqnarray*}
  \hat{\theta}^K \, \hat{a} = (\hat{a} \, A^K_L + \hat{b} \, C^K_L)
     \, \hat{\theta}^L     \quad , \quad
  \hat{\theta}^K \, \hat{b} = (\hat{a} \, B^K_L + \hat{b} \, D^K_L)
     \, \hat{\theta}^L    \quad , \quad
  \hat{\theta}^K \, {\cal D}^{1/2} = \pm (t/q) \, {\cal D}^{1/2}
      \, \hat{\theta}^K
\end{eqnarray*}
and the first two relations with $\hat{a}, \hat{b}$ replaced by
$\hat{c}, \hat{d}$, respectively.
For the plus sign in the last equation the matrices $A,B,C,D$ are
now given by (\ref{ABCD-t+}).
In case of the minus sign they are given by (\ref{ABCD-t-}).
                                  \hfill {\Large $\Box$}
\vskip.3cm
\noindent
{\bf 3.3 The classical limit.}
For the $t_-$ calculus we obtain in particular $\theta^K a =
- a \, \theta^K$ for $K=2,3$ when $q=1$ which is far away from
the ordinary differential calculus on $SL(2)$.
Let us therefore turn to the $t_+$ calculus. For $q=1$ we obtain
\begin{eqnarray}
 \left(\begin{array}{c} \theta^1 \\ \theta^2 \\ \theta^3
                        \\ \theta^4  \end{array}\right)  \, a =
 a \, \left(\begin{array}{c} \theta^1 + {1\over 3} \, \vartheta \\
    \theta^2 \\ \theta^3 \\  \theta^4 - {1\over 3} \, \vartheta
            \end{array}\right) +
 b \, \left(\begin{array}{c} 0 \\ {2 \over 3} \, \vartheta \\ 0 \\ 0
            \end{array}\right)
\end{eqnarray}
and
\begin{eqnarray}
 \left(\begin{array}{c} \theta^1 \\ \theta^2 \\ \theta^3
                        \\ \theta^4  \end{array}\right)  \, b =
 b \, \left(\begin{array}{c} \theta^1 - {1\over 3} \, \vartheta \\
      \theta^2 \\ \theta^3 \\  \theta^4 + {1\over 3} \, \vartheta
            \end{array}\right) +
 a \, \left(\begin{array}{c} 0 \\ 0 \\ {2 \over 3} \, \vartheta \\ 0
            \end{array}\right)
\end{eqnarray}
where now $\vartheta = \theta^1 + \theta^4$.
In terms of $x^\mu$ (see section 2.4), these relations can be expressed
in the form (\ref{x-dx}) with (\ref{tau}) and (\ref{g}) where now
$s = 1/3$ and the differential calculus on $SL(2)$ is {\em not} the
ordinary one.\footnote{One might have expected that we simply had to
insert the value of $t_+$ at $q=1$ in (\ref{s-t}) which would indeed
lead to the {\em ordinary} differential calculus on $SL(2)$. This would
not be correct, however, since {\em before} taking the limit $q \to 1$
we have to identify $t=q$ (rather than treating $t$ as independent of
$q$ as we did at the end of section 2).}
\vskip.2cm

One of the `coordinates' $x^\mu$ is redundant because of the
constraint ${\cal D} = 1$.
Let us consider the subalgebra generated by only three of them,
say $x^i$ where $i=1,2,3$. Then
\begin{eqnarray}
 \lbrack \, x^i \, , \, \mbox{d} x^j \, \rbrack = \tau \, g^{ij}
 \quad , \quad
 g^{ij} = x^i \, x^j + 4 \, \delta^{(i}_2 \, \delta^{j)}_3   \; .
                                                  \label{x-dx-SL}
\end{eqnarray}
Since $\mbox{det}(g^{ij}) = -4 \, (x^1)^2$, $g$ is
a non-degenerate symmetric matrix if $x^1 \neq 0$.
The latter is just the condition allowing us to solve the
determinant constraint for $x^4$. An attempt to express $\tau$ in
the form $\tau = \sum_{i=1}^3 \mbox{d} x^i \, f_i$ with $f_i \in
{\cal A}$ using $x^4 = (1+x^2 x^3)/x^1$ fails. Therefore,
$\mbox{d} x^i$ and $\tau$ are linearly independent in $\Lambda^1
({\cal A})$, regarded as a right $\cal A$-module (see also
\cite{DMH93-Mex}).
\vskip.2cm

Let ${}^\ast$ be an antilinear involution on $\cal A$ (which on
complex numbers acts as complex conjugation).
The reality conditions $(x^\mu)^\ast = x^\mu$ are compatible with the
$SL_q(2)$ commutation relations only when $|q|=1$. These conditions
define the quantum group $SL_q(2,\Rl)$. Assuming the rule
$\, (f \, \mbox{d} h)^\ast = \mbox{d}(h^\ast) \, f^\ast
  \; (\forall \, f,h \in {\cal A})$,
the $t_+$ calculus on $SL_q(2)$ is compatible with the reality
conditions \cite{MH+Reut93}. In the classical limit ($q=1$), we then
have (\ref{x-dx-SL}) with {\em real} functions $x^i$ and a {\em real}
metric $g$ which turns out to be the maximally symmetric Lorentzian
metric on $SL(2,\Rl)$ with negative constant curvature
\cite{MH+Reut93}.

\section{Comments}
\setcounter{equation}{0}
We have obtained a considerable simplification of some of the results
in \cite{MH+Reut93,MH92}. Particular emphasis has been given to
the fact that the classical limit of a bicovariant differential
calculus on a quantum group does {\em not} coincide, in general,
with the ordinary differential calculus.
In particular in view of possible applications of bicovariant
differential calculus on quantum groups in physics, it is interesting
that the resulting `deformed' calculus exhibits relations to
various branches of mathematical physics. This will be discussed
briefly in the following.
A generalization of (\ref{x-dx-SL}) is given by
\begin{eqnarray}        \label{x-dx-g}
  \lbrack \, x^i \, , \, \mbox{d} x^j \, \rbrack = \tau \, g^{ij}
\end{eqnarray}
where $x^i$ are coordinates on a (smooth) manifold $\cal M$, $g$
a contravariant symmetric tensor field (e.g., a metric), and $\tau$
a 1-form on $\cal M$ which commutes with $x^i$, anticommutes with
1-forms and satisfies $\mbox{d} \tau =0$. One can show that
these commutation relations are well defined on $\cal M$, i.e.
independent of the choice of coordinates.
Such a calculus has been considered before
\cite{DMH92-grav,DMH93-stoch} with $\tau = \mbox{dt}$,
the ordinary differential of a real parameter $\mbox{t}$.
As a consequence of (\ref{x-dx-g}) we then have
\begin{eqnarray}
  \mbox{d} f = \mbox{dt} \, \left( {\partial \over \partial \mbox{t}}
  + {1 \over 2} \, g^{ij} \, {\partial \over \partial x^i}
  {\partial \over \partial x^j} \right) \, f
  + \mbox{d} x^i \, {\partial \over \partial x^i} \, f
                      \label{df-t}
\end{eqnarray}
for a function $f(\mbox{t},x^i)$. Note that the differential of
$f$ involves a {\em second order} differential operator. This
hints towards applications of this calculus in the context of
stochastics (diffusion equation), quantum mechanics (Schr\"odinger
equation) and `proper time' relativistic quantum theories (see
\cite{Fanc93} for a review).
For a real (positive definite) tensor field $g$, the first order
calculus was indeed shown to be equivalent to the (It\^o) calculus of
stochastic differentials \cite{DMH93-stoch} (see also \cite{DMH93-Mex}).
Here $\mbox{t}$ is the stochastic time. The other aspects mentioned
above were discussed in \cite{DMH92-grav,MH92-habil}.
\vskip.2cm

In (\ref{x-dx-SL}) the 1-form $\tau$ is {\em not} of the form
$\mbox{dt}$ with a parameter $\mbox{t}$ independent of the $x^i$.
Therefore, there is no (extra) `time' parameter in this case and
we have
\begin{eqnarray}
  \mbox{d} f = \tau \, {1 \over 2} \, g^{ij} \, {\partial \over
               \partial x^i} {\partial \over \partial x^j} \, f
               + \mbox{d} x^i \, {\partial \over \partial x^i} \, f
\end{eqnarray}
instead of (\ref{df-t}).
\vskip.2cm

In this work we have only considered bicovariant differential
calculus on the quantum groups $GL_{p,q}(2), GL_q(2), SL_q(2)$
and $SU_q(2)$. For the corresponding higher-dimensional quantum
groups, one does not have a complete knowledge of the bicovariant
calculi yet (with the exception of $GL_q(3)$ \cite{Bres93} for which
the calculi induced on quantum subgroups are now being studied). But
the existing examples also exhibit a nonstandard classical limit,
in general. This will be discussed in more detail elsewhere.

\section*{Acknowledgment}
 I would like to thank Klaus Bresser for some helpful discussions.


\end{document}